# Precipitation diffusion downscaling and application to out-of-distribution simulations with and without stratospheric aerosol injection

Cameron Dong[1], James W. Hurrell[1], Elizabeth A. Barnes[2,3]

[1]Department of Atmospheric Science, Colorado State University, Fort Collins, CO

[2]Faculty of Computing and Data Sciences, Boston University, Boston, MA

[3]Department of Earth and Environment, Boston University, Boston, MA

*Corresponding author*: Cameron Dong, c.dong@colostate.edu

ABSTRACT

Stratospheric aerosol injection (SAI), a possible climate engineering strategy where reflective particles are injected into the stratosphere, has been explored to mitigate global warming and its associated risks, such as the intensification of extreme precipitation events. However, current Earth system models (ESMs) often used to simulate SAI and other climate change scenarios are too coarse to properly assess such risks. Traditional statistical downscaling methods, used to project higher resolution impacts, may be biased and unrealistic. To address this, we train a deep learning diffusion downscaler to generate 0.25° contiguous United States (CONUS) daily precipitation using historical and future climate simulations from the Mesoscale Atmosphere-Ocean Interaction in Seasonal-to-Decadal Climate Prediction (MESACLIP) project, then apply the diffusion downscaler to out-of-distribution CESM2 simulations with and without SAI. The diffusion model generates realistic downscaled precipitation using either MESACLIP or CESM2 inputs. It also faithfully recreates the climate change projections of extreme precipitation in MESACLIP. Diffusion-downscaled projections of the future CESM2 SAI scenarios suggest that SAI could nearly cut in half the CONUS-average increase in yearly max precipitation, compared to the non-SAI scenario. However, there is considerable regional variation and internal variability, with SAI modeled to only slightly reduce increases in extreme precipitation frequency in the Mid Atlantic and the Pacific Northwest, but mitigating most intensification in other regions. Future application of diffusion downscaling to a wider variety of SAI scenarios would provide valuable insight into how proposed SAI strategies may affect precipitation variability on fine spatial scales for regional impact assessments.

## 1. Introduction

Most projections of climate change and its impacts rely on future climate simulations using Earth system models (ESMs), which are used to explore a wide variety of hypothetical future climate scenarios. Considering current carbon dioxide emissions mitigation efforts, the most likely future scenarios will result in warming exceeding the 1.5°C Paris temperature target (Matthews & Wynes, 2022; Rogelj et al., 2016). One proposed idea to still achieve this target is stratospheric aerosol injection (SAI), a potential climate intervention strategy to alleviate some impacts of global warming by injecting reflective particles into the stratosphere that redirect incoming sunlight and cool the planet (National Academy of Sciences et al., 2021; Crutzen 2006; Hulme, 2012). SAI, compared to other climate intervention technologies (e.g. carbon dioxide removal), is considered more technologically feasible while being relatively cheap to implement (Robock et al., 2009; Smith & Wagner, 2018). However, while it is simulated to effectively mitigate global rises in temperature (MacMartin et al., 2018; Tilmes et al., 2020), it may not directly combat other negative effects of rising carbon dioxide. Additionally, there may be different (and potentially harmful) regional impacts, compared to a scenario without any intervention (Halstead, 2018; Hueholt et al., 2024; Irvine et al., 2017; Tang & Kemp, 2021).

To-date, understanding the potential positive or negative impacts of SAI relies largely upon ESM simulations. However, properly assessing and planning for such impacts requires information on spatial scales that are too small or too expensive to resolve with most Earth system models (Maraun 2016, Marotzke et al., 2017). ESMs typically run at coarse resolutions (grid size >= 1°) that are computationally cheaper and allow for a wider exploration of scenario spaces and internal climate variability, but may lead to inaccurate representations of many climate variables, particularly when analyzing extremes on regional and local scales (Kopparla et al., 2013; Iles et al., 2020; Sillmann et al., 2017). This is especially true for precipitation, which exhibits strong variations in model representation of orography and convection at different resolutions, hindering representations of heavy rainfall, flooding, and other precipitation-related impacts that depend on spatial scale.

Here we consider "high-resolution" as having grid spacing less than or equal to 0.25° (in contrast to "km-scale" with grid spacing less than 10 km). Some high-resolution global climate simulations do exist for the most common future climate scenarios (e.g. HighResMIP, Haarsma et al., 2016), such as the Shared Socioeconomic Pathways (SSPs; Riahi et al., 2017). However, such simulations for SAI scenarios are much less common (Feder et al., 2026). Partially, this is because of the high computational cost of SAI simulations (which may need both complex stratospheric chemistry and increased vertical resolution). An alternative to running high-resolution SAI climate simulations is to downscale existing coarse climate simulations using dynamical or statistical methods. Dynamical methods typically target a specific region, using a high-resolution regional climate model, with boundary forcing provided by the low resolution ESM data (Giorgi and Gutowski 2015; Feser et al., 2011). Such dynamical methods are also computationally expensive, especially at kilometer-scales, and additionally are subject to the regional climate model and forcing ESM biases (Xu et al., 2019). Statistical methods consist of determining empirical relationships between biased coarse ESM data and the target high-resolution data (Wilby and Wigley 1997; Maraun and Widmann, 2018). A benefit of statistical methods is their much cheaper computational cost and simplicity. However, traditional statistical methods are often linear and may generate samples that are less physically realistic and may not represent extremes well (Chandel et al., 2024, Gutmann et al., 2014).

As an alternative, machine learning approaches have recently been developed to perform downscaling with greater fidelity than traditional statistical methods (Rampal et al., 2024). In general, these methods use machine learning algorithms to learn the relationships between coarse predictors and fine scale variability in a data-driven manner. In the past few years, deep learning methods (LeCun et al., 2015) have been developed that can outperform traditional downscaling methods (Baño-Medina et al., 2020; Hobeichi et al., 2023, Rampal et al., 2022). In particular, downscaling has been performed using diffusion modeling, a probabilistic deep learning framework that can stochastically generate high-resolution images given low-resolution inputs (Ho et al., 2020; Song et al., 2020). A strength of diffusion models is their ability to faithfully reproduce the target data distribution and generate realistic samples, compared to typical deterministic machine learning methods. Previous research has shown diffusion models perform well at representing extremes in downscaling situations, which is important for assessing climate risks such as flooding and heat waves (Addison et al., 2026; Lopez-Gomez et al., 2025; Mardani et al., 2025; Aich et al., 2026).

While there is some research applying dynamical and traditional downscaling methods to SAI simulations (Wang et al., 2022; Wang et al., 2024), only recently have machine learning methods been applied for monthly data (Wang et al., 2026). Furthermore, to our knowledge deep learning methods have not yet been used to downscale projections of extreme precipitation in SAI future scenarios. To help fill this gap, we develop a diffusion downscaler to generate daily high-resolution (0.25°) precipitation over the contiguous United States (CONUS). The downscaler is trained on high-resolution historical and future global warming simulations performed using the Community Earth System Model version 1 (CESM1) from the Mesoscale Atmosphere-Ocean Interaction in Seasonal-to-Decadal Climate Prediction (MESACLIP) project (Chang et al., 2020). We then apply it to coarse simulations from CESM2 with and without SAI, in order to assess future impacts to extreme precipitation related to the climate intervention strategy.

It is important to note that the training data (from CESM1) and application SAI data (from CESM2) are run with different model versions that have different variability and biases due to changes in numerical schemes, physical parameterizations, and resolution (Simpson et al., 2020). Previous research has found that downscaling performance may degrade when the model is fed inputs outside of its training distribution (Aich et al., 2026; Wan et al., 2026; Wang and Tian 2024). As such, before applying the diffusion downscaler to future CESM2 simulations with and without SAI, we test the out-of-distribution performance of the machine learning model over a historical calibration period. We also evaluate whether the downscaler successfully reproduces the climate change signal in the MESACLIP test data. Having validated the performance of the diffusion downscaler, we assess the differences in extreme precipitation under an SAI scenario that limits global temperature to the Paris 1.5° warming target, compared to a moderate emissions scenario with no climate intervention.

## 2. Methods and Data

### *2.1 MESACLIP simulation data*

We train and test our machine learning downscaler using high-resolution climate simulations from MESACLIP (Chang et al., 2020). MESACLIP uses the Community Earth System Model version 1.3 (CESM1.3; Meehl et al., 2019) with the Community Atmosphere Model version 5 (CAM5; Neale et al., 2012) run at ~0.25°. The model is low-top, with only coarse resolution in the stratosphere and totaling 30 vertical levels up to 3 hPa. This high-resolution version of CESM1.3 is much improved compared to the low-resolution version (~1°) in representing extremes, such as those related to tropical cyclones, atmospheric rivers, and extreme precipitation over CONUS (Chang et al., 2020; Chang et al., 2025).

MESACLIP includes a variety of ensemble experiments using different forcing scenarios. In particular, we train, validate, and test our model on simulations with historical forcing (from 1920-2005) and under representative concentration pathway 8.5 (RCP8.5; 2006-2070). Training under both a historical and business-as-usual warming scenario ensures that the neural network is able to predict extremes under a variety of warming levels, not having to extrapolate dynamical relationships from the present period into the future (Rampal et al., 2024b). This ensures that the machine learning model primarily needs to generalize to the biased CESM2 inputs, rather than additionally extrapolating potentially non-stationary relationships into an unseen warmer climate.

The requisite daily data is available from 9 ensemble members for each set of forcings. The daily output used includes geopotential height at 500 hPa (Z500), total column water (TCW), and mean precipitation rate. All data is regridded bilinearly from the CESM1.3 native grid to a regular 0.25° grid.

### *2.2 CESM2 climate simulations with and without SAI*

After training the diffusion downscaling model (see section 2.3) on MESACLIP historical and future simulations, we apply the downscaler to coarse output from existing coupled climate simulations with and without SAI. These simulations have been performed using CESM2 (Danabasoglu et al., 2020) with the high-top atmospheric component Whole Atmosphere Community Climate Model version 6 (WACCM6; Gettelman et al., 2019a). The nominal horizontal resolution is ~1°, with 70 vertical layers and a model top at $4.5\times10^{-6}$ hPa, necessary to represent various stratospheric dynamical and chemical processes, such as those associated with SAI. The simulations consist of historical simulations run as part of the Coupled Model Intercomparison Project Phase 6 (CMIP6; Eyring et al., 2016) from 1850 to 2014, with two varying future scenarios.

The first future simulations use an SSP2-4.5 scenario (Riahi et al., 2017; Meehl et al., 2020), which is a middle-of-the-road scenario with moderate mitigation, compared to more extreme business-as-usual scenarios (e.g. RCP 8.5 used to train the downscaler). The second future scenario consists of corresponding simulations branched at year 2035 from the SSP2-4.5 simulations, maintaining the same emissions trajectory but where SAI is used to restrict global temperatures to 1.5° of warming relative to the preindustrial period, referred to as Assessing Responses and Impacts of Solar climate intervention on the Earth system with Stratospheric Aerosol Injection (ARISE-SAI; Richter et al., 2022). These simulations use an optimized controller algorithm to manage stratospheric injection rates across various latitudes to additionally reduce changes in the mean spatial pattern of temperature (MacMartin et al., 2014; Kravitz et al., 2017).

Although ten ensemble members of data are available for each future simulation type, only three members are available for the historical period, so we restrict our analysis to three members for all simulations. While the SSP2-4.5 simulations run until the year 2100, the ARISE-SAI simulations only run through the year 2069.

*2.3 Diffusion Downscaling*

We produce downscaled predictions of daily 0.25° precipitation over CONUS (20°N-52°N, 232°E-296°E) using a residual conditional diffusion framework. As conditioning we use coarse precipitation as well as TCW and Z500, which allow the neural network to learn connections between large-scale dynamical conditions and the target precipitation field. The diffusion model predicts the fine-scale residual difference from the coarse conditioning precipitation. The predicted residual and input coarse precipitation are then summed together to provide the final downscaled field.

All conditioning variables are conservatively regridded onto a 4° grid as a preprocessing step, which means that precipitation is technically being downscaled from 4° to 0.25°. While it is possible to downscale directly from 1° (the native CESM2 resolution), 4° is closer to the effective resolution of typical 1° climate simulations (Klaver et al., 2020). Other work has found better generalization capability when conditioning inputs from both the training data (e.g. MESACLIP) and application data (e.g. CESM2) have similar spatial power spectral densities (Aich et al., 2026). Similarly, we have found that the model generalizes poorly to CESM2 applications when trained with 1° inputs (not shown).

The 4° predictors are bilinearly interpolated back onto the 0.25° grid when downsampling, which maintains only the coarse spatial structure of the variables while allowing them to be used as input for the neural network architecture. It is important to emphasize that our goal is *not* to generate highest "skill" downscaler which produces the lowest possible error on MESACLIP, but instead to maximize the ability of the downscaler to generate *realistic* variability and climate change signals on both MESACLIP and the out-of-distribution CESM2 inputs.

Coarse predictors are normalized using their mean and standard deviation across all grid points and timesteps from MESACLIP training ensemble member 002 of both the historical and RCP8.5 period. Precipitation is predicted in units of mm/hour and transformed using the square root, which was found to provide a reasonable scaling for the neural network training.

Figure 1 depicts the diffusion downscaling process. Daily inputs of the aforementioned coarse predictors, along with high-resolution 0.25° invariant inputs (surface geopotential, land fraction, latitude, longitude) are provided to the neural network as conditioning. Random gaussian noise is sampled, then iteratively transformed by the neural network into a map of high-resolution daily rainfall. The process is probabilistic in the sense that many samples of high-resolution precipitation can be produced for the same coarse predictors by starting the denoising process from independent gaussian samples.

The neural network is trained using the flow matching objective (Lipman et al., 2023), with uniform sampling of noise during training. The neural network consists of a standard U-net with residual blocks within each individual layer and skip connections between the encoding and decoding layers of the same horizontal dimensions, with sinusoidal time-embeddings to manage the noise level (Figure S1). We test a few hyperparameters, which consist of initial learning rate (1e-4, 2e-5) and batch size (16, 64). All training follows a cosine schedule for the learning rate with a maximum epoch size of 25. The coarse inputs (precipitation, TCW, Z500) are collectively dropped during training with a probability of 0.1 and filled with 0's. During inference, classifier-free guidance can be used with different strengths (-0.1, 0, 0.1), for which the optimal strength is determined using the validation dataset. Exponential decay moving average is used for the model weights, with the momentum factor decreasing with batch size (so that approximate momentum per epoch is equivalent). The best model training hyperparameters are selected based on the lowest validation loss (i.e. mean squared error from the flow matching objective). For the selected training hyperparameters, the optimal guidance strength and checkpoint are chosen based on which best reproduces the probability distribution skill score over the historical period in the validation ensemble member (Perkins et al., 2007). Using one Nvidia H100 GPU, training a single model requires 12 hours. Generating samples for one ensemble member of the 25-year test period (future or historical) requires 5 minutes. This is a significant speedup over running even the coarse-resolution ESM.

Training data consists of 5 ensemble members each of the MESACLIP historical and RCP8.5 simulations, each from 1920-2070, yielding 750 years of training data. Validation data consists of 1 ensemble member during the historical period 1980-2004. Testing data consists of 3 ensemble members, during the 1980-2004 historical and 2045-2069 future periods. We apply the diffusion downscaling model to inputs from CESM2 during the same years for the historical and future scenarios, respectively. The CESM2 historical and future inputs are bias-corrected on the 4° grid with quantile delta mapping (QDM) before being used as input (Cannon et al., 2015), utilizing the MESACLIP historical simulations for calibration. Such bias-correction corrects the distribution at each grid point (but not the temporal evolution or spatial and intervariable relationships), which we have found necessary to improve the generalization capability of the diffusion downscaler to CESM2 inputs.

While multiple unique samples can be generated using the diffusion model for the same coarse inputs on any individual day, we only generate one sample per day throughout the analyses, unless otherwise specified. This allows for an equal comparison with the existing MESACLIP data. For applications where generating more samples per day would be useful, the diffusion downscaler generates a well-calibrated spread of values, according to a rank histogram analysis (Figure S2).

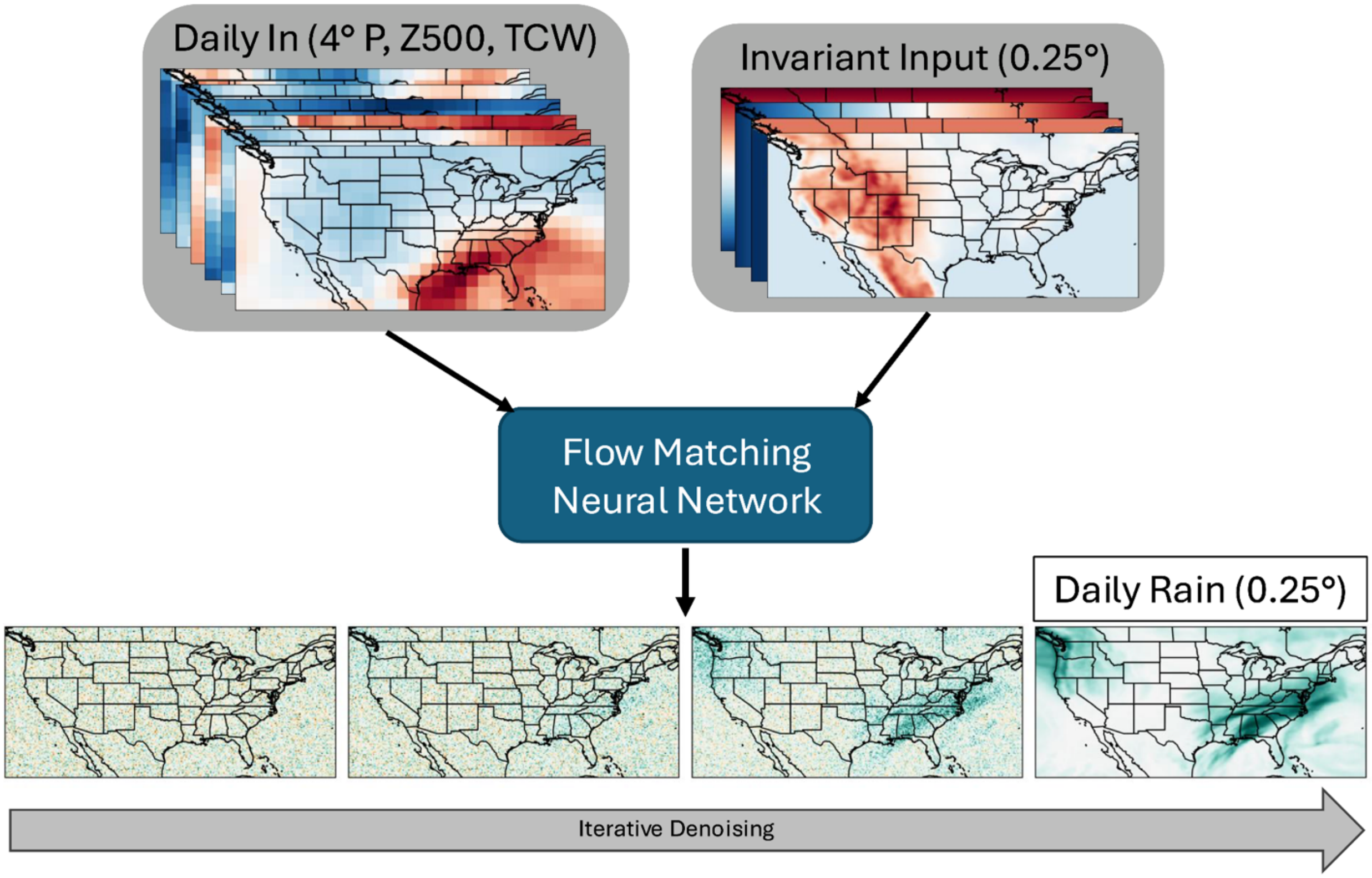

**Figure 1:** *Schematic of the process to produce downscaled daily precipitation, using coarse daily input as a guide to the diffusion neural network to perform iterative denoising.*

### *2.4 Statistical Downscaling Baseline*

To provide a baseline downscaler for comparison to our diffusion approach, we implement a daily version of the Bias Correction with Spatial Disaggregation (BCSD) method (Thrasher et al., 2012), which only uses coarse precipitation as input. Daily BCSD has been commonly utilized to evaluate diffusion downscaling skill (e.g. Addison et al., 2026; Lopez-Gomez et al., 2025).

When used with MESACLIP to compare BCSD with the diffusion downscaler, we first conservatively remap MESACLIP precipitation to the native CESM2 grid (0.9° by 1.25°), then spatially disaggregate, in order to mimic how BCSD would be used with actual CESM2 data in real world applications. When using CESM2 precipitation as input to BCSD (e.g. section 3.3), we bias-correct the native resolution CESM2 precipitation to match MESACLIP with QDM before performing the spatial disaggregation step.

## 3. Results

### *3.1 Evaluation over historical period*

We begin by comparing the variability of downscaled precipitation from the diffusion model with actual high-res MESACLIP precipitation during the historical period (1980-2004), when conditioned on either MESACLIP or CESM2 coarse inputs. In order to later apply the diffusion model to future CESM2 climate scenarios (section 3.3), it is important to first verify that the CESM2-conditioned neural network produces precipitation variability similar to the original MESACLIP over the historical period with comparable external forcings (since there is no "true" CESM2 high-resolution precipitation to compare with). The diffusion outputs are contrasted with the daily BCSD method as a traditional statistical downscaling baseline.

The diffusion model is capable of capturing the statistics of precipitation over CONUS. It closely reproduces the mean precipitation from MESACLIP, when conditioned on MESACLIP or CESM2 coarse predictors (Figure 2c,e). While the differences from the true MESACLIP precipitation are slightly larger for CESM2 conditioning, they are still less than 5% on average across CONUS. The small errors likely stem from biases in the spatial patterns of the coarse CESM2 predictors, despite regridding to a more well-resolved grid spacing and the usage of QDM. By design, BCSD nearly matches the climatological precipitation rates in MESACLIP, although the extremes are significantly underestimated as will be discussed next (Figure 2g).

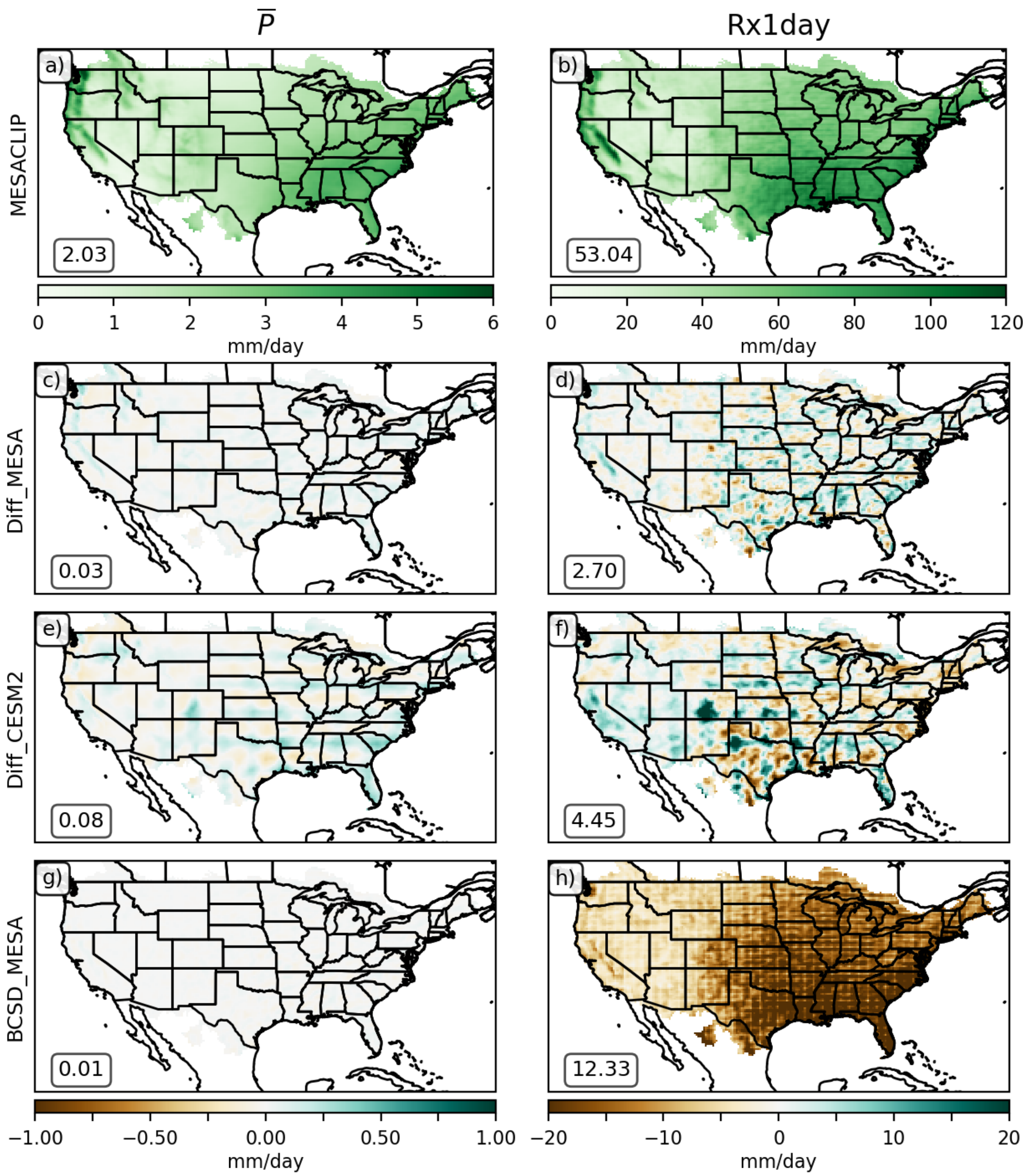

**Figure 2:** *(top) MESACLIP climatological mean precipitation (left) and climatological Rx1day precipitation (right) over CONUS. Differences from true MESACLIP in diffusion-downscaled MESACLIP (Diff_MESA, middle upper), diffusion-downscaled CESM2 (Diff_CESM2, middle lower) and BCSD_MESA (lower). Values in bottom left corner of each subpanel show the mean absolute value over CONUS.*

The picture is slightly different for extreme precipitation (here represented by yearly max precipitation; Rx1day) which is more difficult for traditional statistical downscaling methods to reproduce. The MESACLIP-conditioned diffusion model produces slightly too extreme values on the west coast over the mountainous areas (Figure 2d). Over the central and eastern United States, there is no clear dry or wet bias, with errors only occurring on very small spatial scales, due to the stochastic nature of precipitation extremes. When conditioned on CESM2 inputs, the diffusion model produces slightly larger, more spatially cohesive differences from MESACLIP (Figure 2f). In general, there are slightly wetter extremes over the southwest United States, with slightly drier extremes over the Northeastern United States. This is likely the result of biases in the spatial patterns and intervariable relationships in CESM2, which cannot be corrected with traditional bias-correction methods like QDM.

In any case, CESM2-conditioned diffusion still produces more realistic results than BCSD, which produces too weak Rx1day precipitation. This is because the spatial disaggregation step in BCSD spreads precipitation over many small grid cells, particularly in regions with spatially homogenous mean precipitation fields, such as the Eastern United States, where BCSD misses the wettest extremes. Meanwhile, the diffusion downscaler can stochastically place precipitation, resulting in more realistic, larger spatial variance and extremes at small scales. This is evident when analyzing individual samples of precipitation, for which the diffusion model generates appropriate spectral power across nearly all wavelengths (Figure S3). Meanwhile, daily BCSD smooths data at medium and short wavelengths, resulting in much lower spectral power (as expected from the spatial disaggregation method).

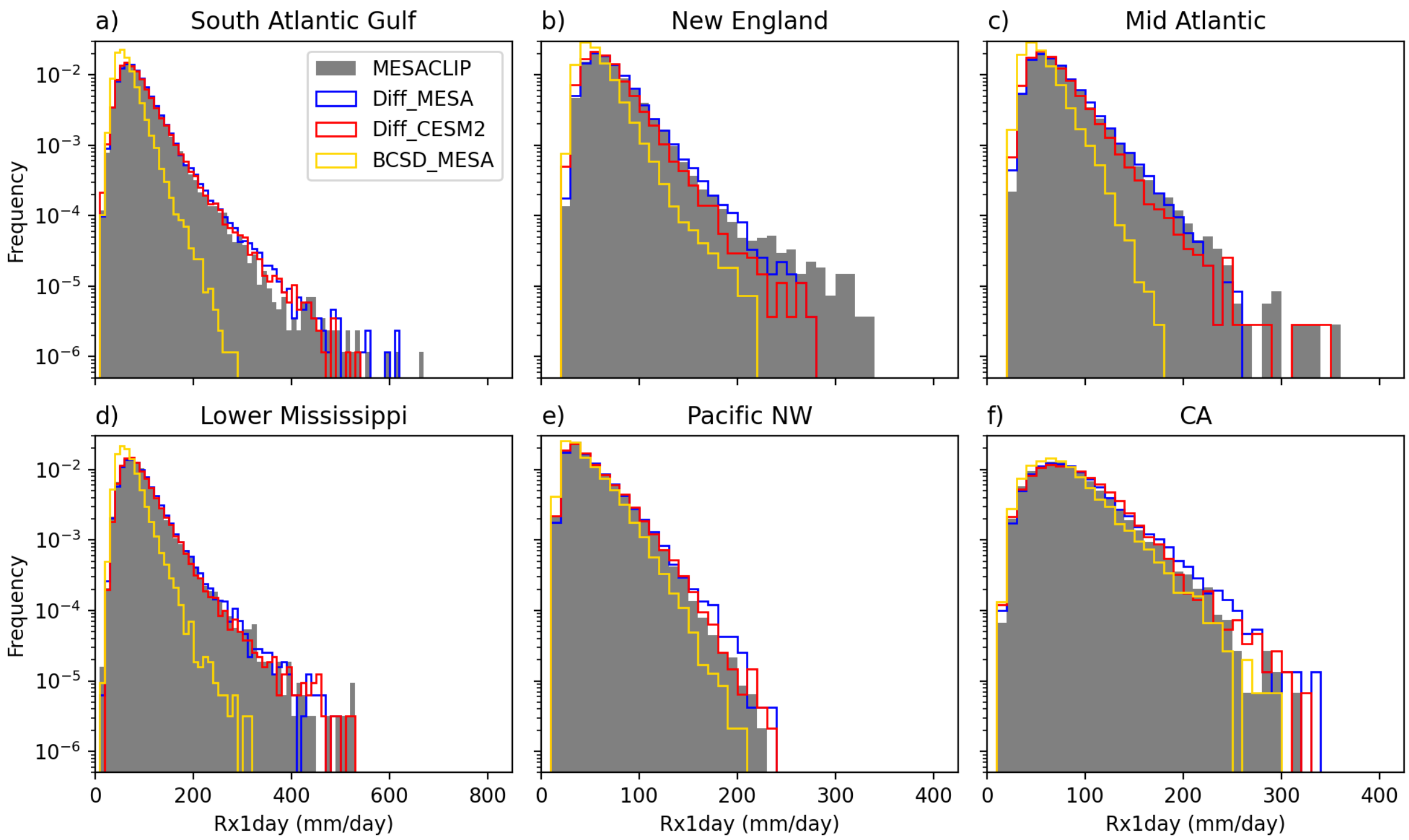

**Figure 3:** *Normalized probability distribution of Rx1day values during the historical period and across grid cells in the (a) South Atlantic Gulf, (b) New England, (c) the Mid Atlantic, (d) the Lower Mississippi, (e) the Pacific Northwest, and (f) California watersheds. Gray shading represents actual MESACLIP values, while blue and red represent diffusion downscaled MESACLIP (Diff_MESA) and CESM2 (Diff_CESM2) respectively. Yellow represents downscaled MESACLIP values using daily BCSD (BCSD_MESA). Bin width is 10 mm/day. Only historical climatologically wet grid cells (greater than 2 mm/day) are included.*

A more complete picture is obtained by analyzing the distribution of Rx1day values (Figure 3), focusing on the aggregate distributions from the wettest United States Geological Survey watersheds along with west and east coasts of CONUS (U.S. Geological Survey 2025). It is important to capture the full intensity of these most extreme events, which have disproportionate ramifications for flooding, runoff and nutrient availability, groundwater storage, and human health and infrastructure (Gimeno et al., 2022; Chang et al., 2023; Zhang et al., 2016).

The diffusion model performs well at capturing the most extreme rainfall rates across various geographic regions, whether conditioned by MESACLIP or CESM2 coarse inputs. While conditioning on CESM2 does lead to slightly weaker extremes in New England and the Mid Atlantic (Figure 3b,c), it is still a much better representation than BCSD, which produces maximum precipitation rates that are only about half of those present in MESACLIP across the eastern United States. There is less of a difference over the west coast watersheds (Pacific Northwest and California; Figure 3e,f), likely because of the strong orographic role in precipitation there. These results highlight the ability of the diffusion model to learn the relationships between coarse inputs and fine-scale extreme precipitation across a wide variety of geographic regions, with differing physical drivers of precipitation, in an entirely data-driven manner.

While it is important to capture extreme precipitation events, it is also critical to accurately represent the full distribution of precipitation values, from low to high intensities. For example, hydrological and land models are highly sensitive to the distribution of precipitation values and length of wet spells (Chen et al., 2013). We quantify how well the diffusion model reproduces the MESACLIP precipitation distribution through a quantile-quantile analysis (Figure 4). Aggregated within each watershed, we calculate the 1st through 99.95th (~2000 day return period) quantile of precipitation values for each of the downscaling methods and compare the results with the true MESACLIP precipitation.

Conditioned on MESACLIP, the diffusion downscaler closely matches all quantiles of precipitation, with only a slight positive bias (1-3%) for both moderate and heavy rainfall quantiles (Figure S4). The exception is for the most extreme precipitation quantile in California, which is about 5% too wet for the MESACLIP-conditioned diffusion downscaler.

The diffusion downscaler also performs well when conditioned on CESM2 inputs (Figure 4a,d,e; Figure S4a,d,e). In the eastern United States watersheds, the CESM2-conditioned diffusion downscaler slightly overestimates moderate precipitation, while it slightly underestimates extreme precipitation in New England and the Mid Atlantic. BCSD also produces an overestimation of moderate rain and underestimation of extreme rain, although across all basins and to a larger extent than the diffusion model. Overall, this indicates that the diffusion model generalizes well to inputs from CESM2, skillfully reproducing both extreme and moderate precipitation during the historical MESACLIP calibration period.

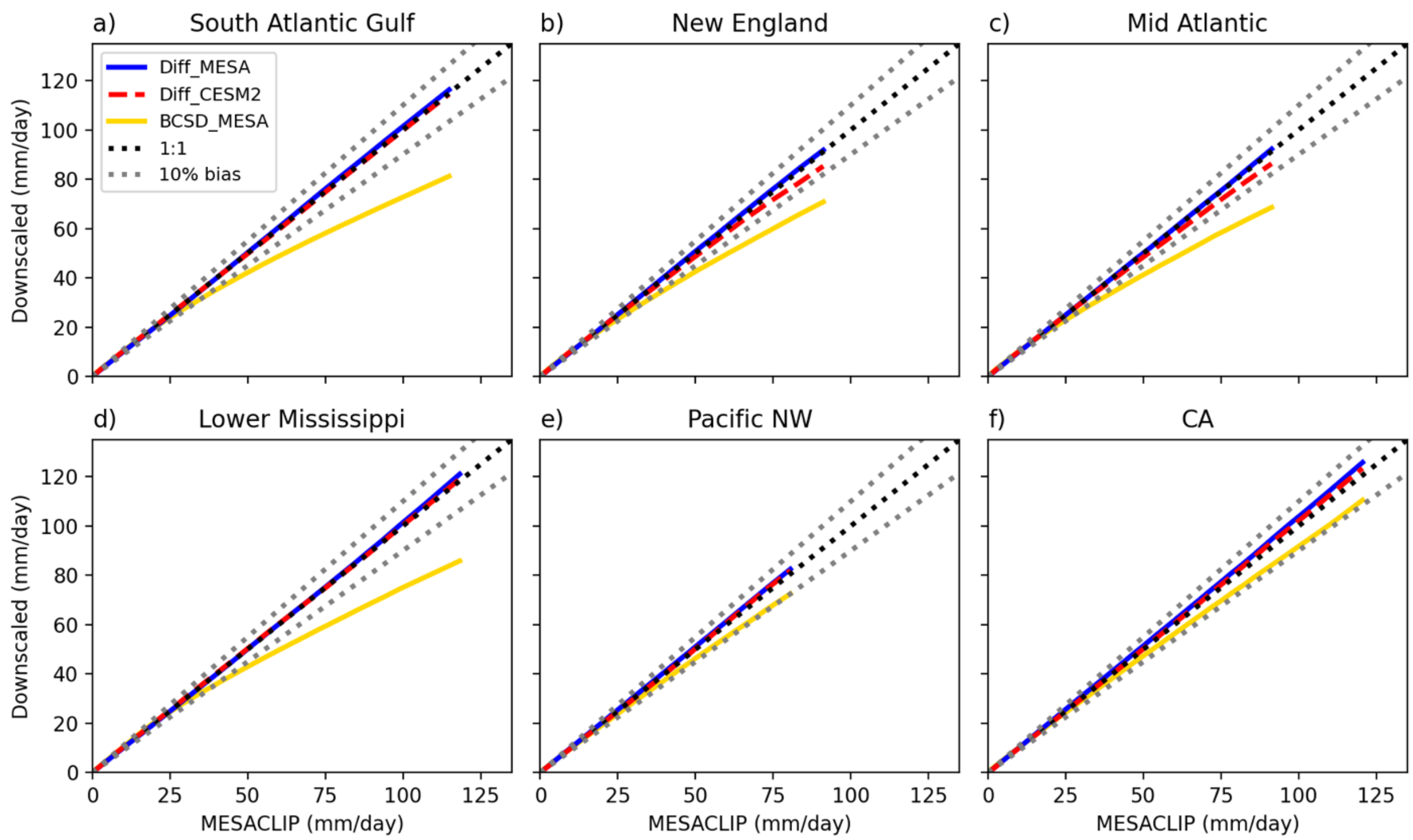

**Figure 4:** *Quantile-quantile plots of precipitation values during the historical period and aggregated across grid cells in the (a) South Atlantic Gulf, (b) New England, (c) Mid Atlantic, (d) Lower Mississippi, (e) Pacific Northwest, and (f) California watersheds. Black dashed line represents 1:1 line (perfect downscaled quantiles). Blue and red lines represent diffusion downscaled MESACLIP (Diff_MESA) and CESM2 (Diff_CESM2), respectively. Yellow represents downscaled MESACLIP values using daily BCSD (BCSD_MESA). Includes quantiles ranging from 0.01 to 0.9995. Only historical climatologically wet grid cells (greater than 2 mm/day) are included.*

### *3.2 Evaluation of MESACLIP climate change signal*

So far, we have shown that the diffusion model accurately reproduces the distribution of precipitation in the historical climate period, including wet extremes, across various geographic watersheds in CONUS. This is true whether the diffusion model is conditioned on MESACLIP or CESM2, although there is slight degradation in certain regions when using biased CESM2 inputs. Now, we validate how well the diffusion model captures the climate change signal. Since there is no “truth” for CESM2 validation in this case, we focus on MESACLIP. This is because the climate change signal is expected to be different in MESACLIP and CESM2 due to different climate scenarios and model physics. The climate change signal under the CESM2 future scenarios with and without SAI is explored in section 3.3.

Figure 5 displays the change in mean and Rx1day precipitation for MESACLIP, the diffusion model, and BCSD, between the historical period (1980-2004) and future RCP8.5 period (2045-2069). For mean precipitation rates, both the diffusion model and BCSD are nearly identical to the true MESACLIP signal. For yearly max precipitation, BCSD slightly underestimates the signal, with a CONUS-averaged increase of 4.13 mm/day. By comparison, the diffusion model closely matches the CONUS-averaged increase (5.14 versus 5.26 mm/day), and it faithfully reproduces the shift in the full distribution of Rx1day values across various watersheds (Figure S5).

Both BCSD and the diffusion model match the large-scale pattern in extreme precipitation changes, with only slight increases in extremes over the west coast, but large increases over much of the Eastern United States. However, the diffusion model and MESACLIP show more fine scale noise (Figure 5b,d), similar to what occurs for individual precipitation samples (Figure S3). In general, BCSD considerably underestimates internal variability on small spatial scales, while the diffusion model closely matches MESACLIP (Figure S6). As a result, projections from the diffusion downscaler are much more appropriate for analyzing uncertainty in the future climate changes in extreme precipitation on fine scales.

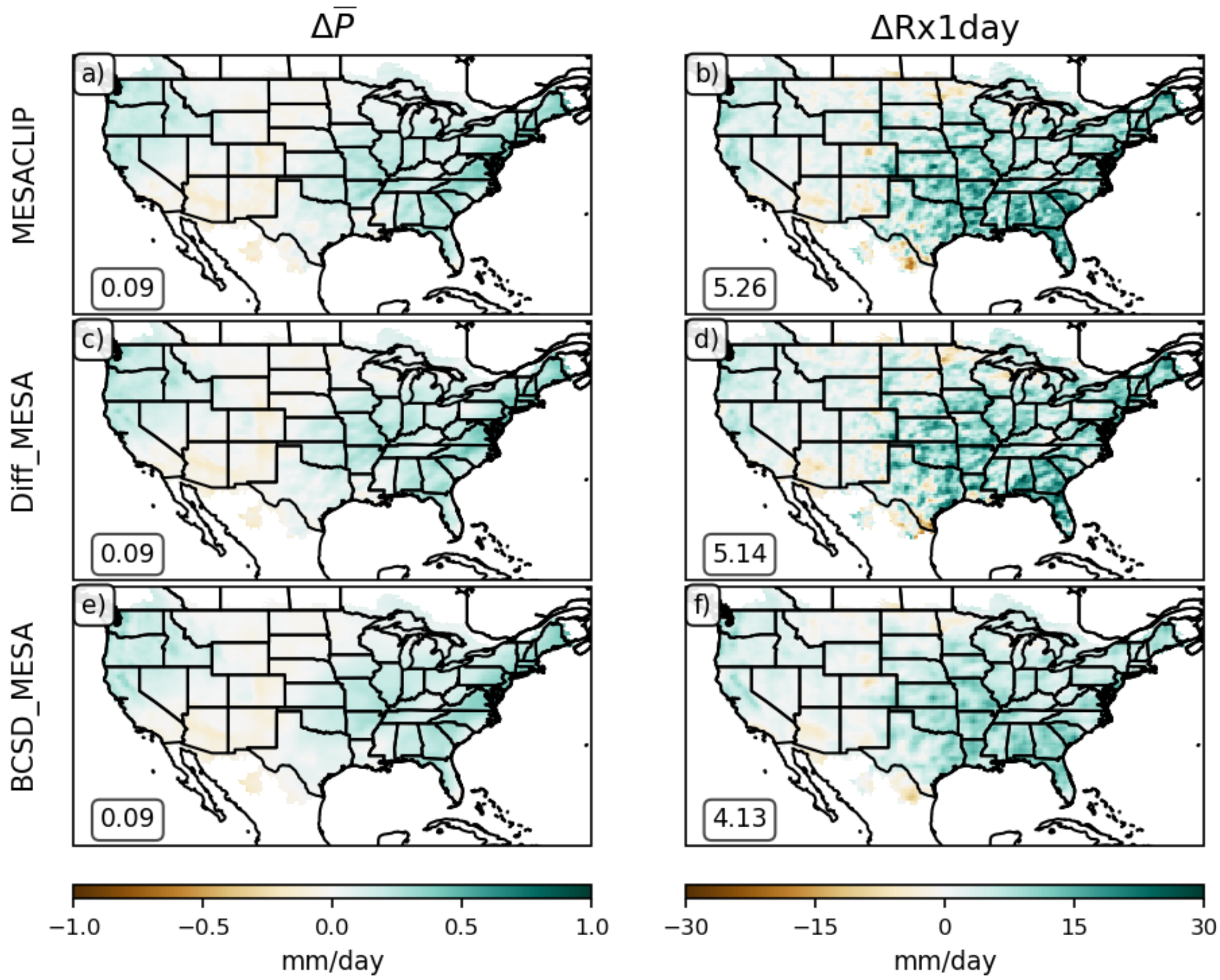

**Figure 5:** *Change in mean (upper) and Rx1day (lower) precipitation between the historical and RCP 8.5 periods. Values from MESACLIP (left), diffusion-downscaled MESACLIP (Diff_MESA, middle) and BCSD_MESA (right). Values in bottom-left corners indicate the CONUS-average value for each subpanel.*

Although the diffusion model realistically reproduces spatial patterns of change in both mean and Rx1day precipitation, it is important to also capture changes in the distribution of precipitation. To address this, we calculate the percent change in each quantile of precipitation, aggregated across grid cells within each watershed of interest for MESACLIP, BCSD, and the diffusion downscaler (Figure 6).

It is notable that except in California and the Pacific Northwest, the largest percentage changes occur for extreme precipitation values, with relatively smaller percentage changes occurring for moderate precipitation quantiles, in agreement with prior research of precipitation changes in a warming climate (O'Gorman and Schneider, 2009; Pendergrass et al., 2017). For example, in the South Atlantic Gulf, moderate precipitation quantiles increase by about 5-6% (80th to 90th percentiles), whereas the most extreme values increase by more than 15% (Figure 6a). Across the eastern US watersheds, the lowest increase in precipitation occurs around the 95th quantile, even decreasing in the South Atlantic Gulf and the Lower Mississippi.

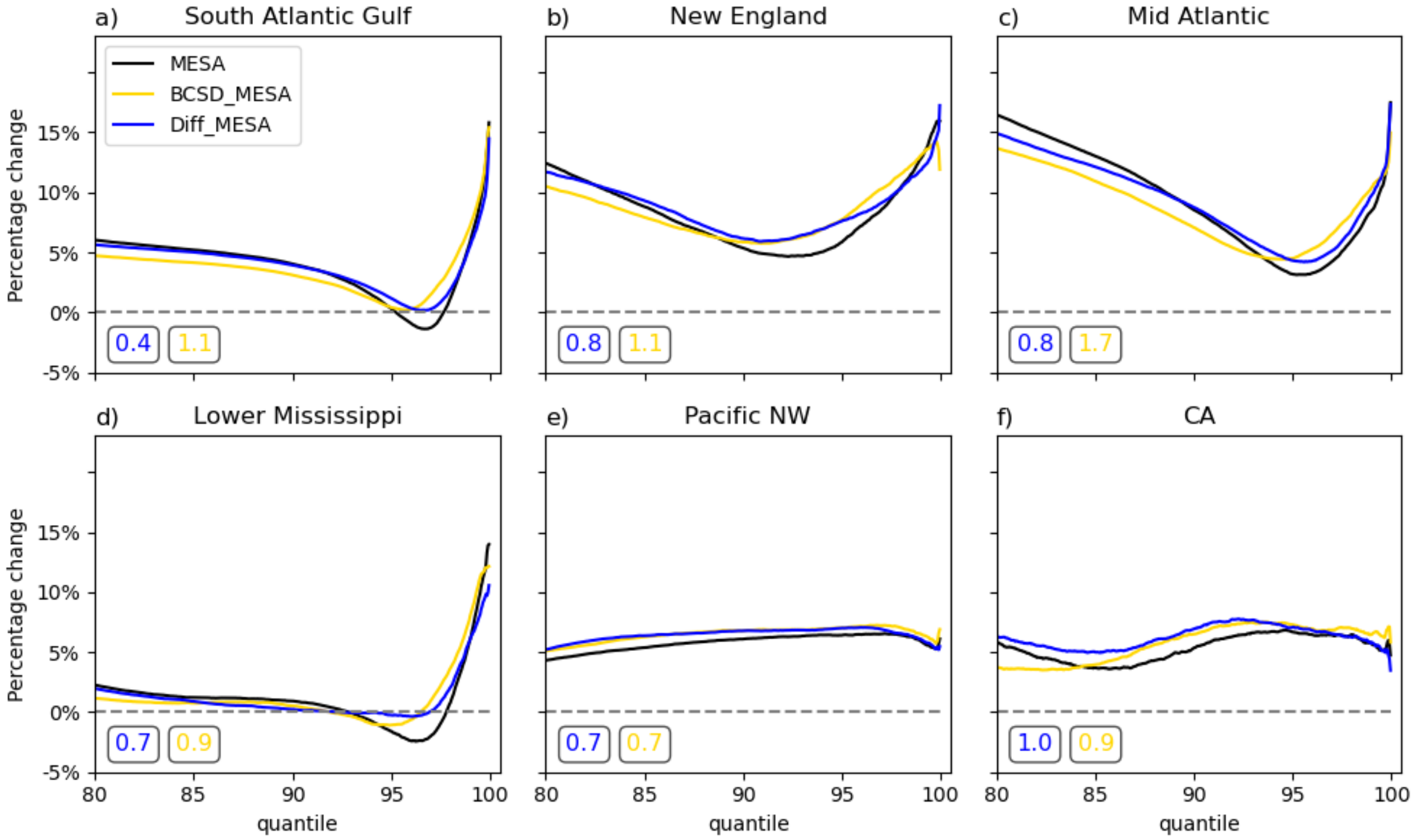

**Figure 6:** *Percent change in daily precipitation rates at various quantiles between the historical period and the future RCP 8.5 period in the (a) South Atlantic Gulf, (b) New England, (c) Mid Atlantic, (d) Lower Mississippi, (e) Pacific Northwest, and (f) California watersheds. Quantiles are calculated using aggregated values from only historically climatologically wet grid cells (greater than 2 mm/day) in each watershed. Numbers in upper-right represent mean absolute difference from the true MESACLIP climate signal for diffusion-downscaled MESACLIP (Diff_MESA, blue) and daily BCSD_MESA (yellow).*

In general, the diffusion model performs slightly better than BCSD at capturing this changing strength of the climate change signal with precipitation intensity, although the pattern is similar using either downscaling method. Relatedly, there is a similar relationship between the frequency of extreme events and intensity, as shown by Figure 7, which displays the global warming-induced change in frequency of extreme precipitation thresholds from the historical period. For example, across the various watersheds, there is very little change in the frequency of the historical 95th precipitation percentile. However, in MESACLIP there is approximately an 80% increase in the frequency of the 99.95th percentile of precipitation in the Eastern United States watersheds.

Both the diffusion model and BCSD accurately represent the change in frequency of different precipitation extreme values in each watershed. Overall, while the diffusion model is better for projecting changes in the raw values of precipitation extremes and capturing the spatial variability and uncertainty associated with extremes, both the diffusion model and BCSD are similar in this case for capturing the percent changes averaged within each watershed. However, when applying both methods to coarse output from CESM2 (section 3.3), it will become apparent that the diffusion model and BCSD produce considerably different results.

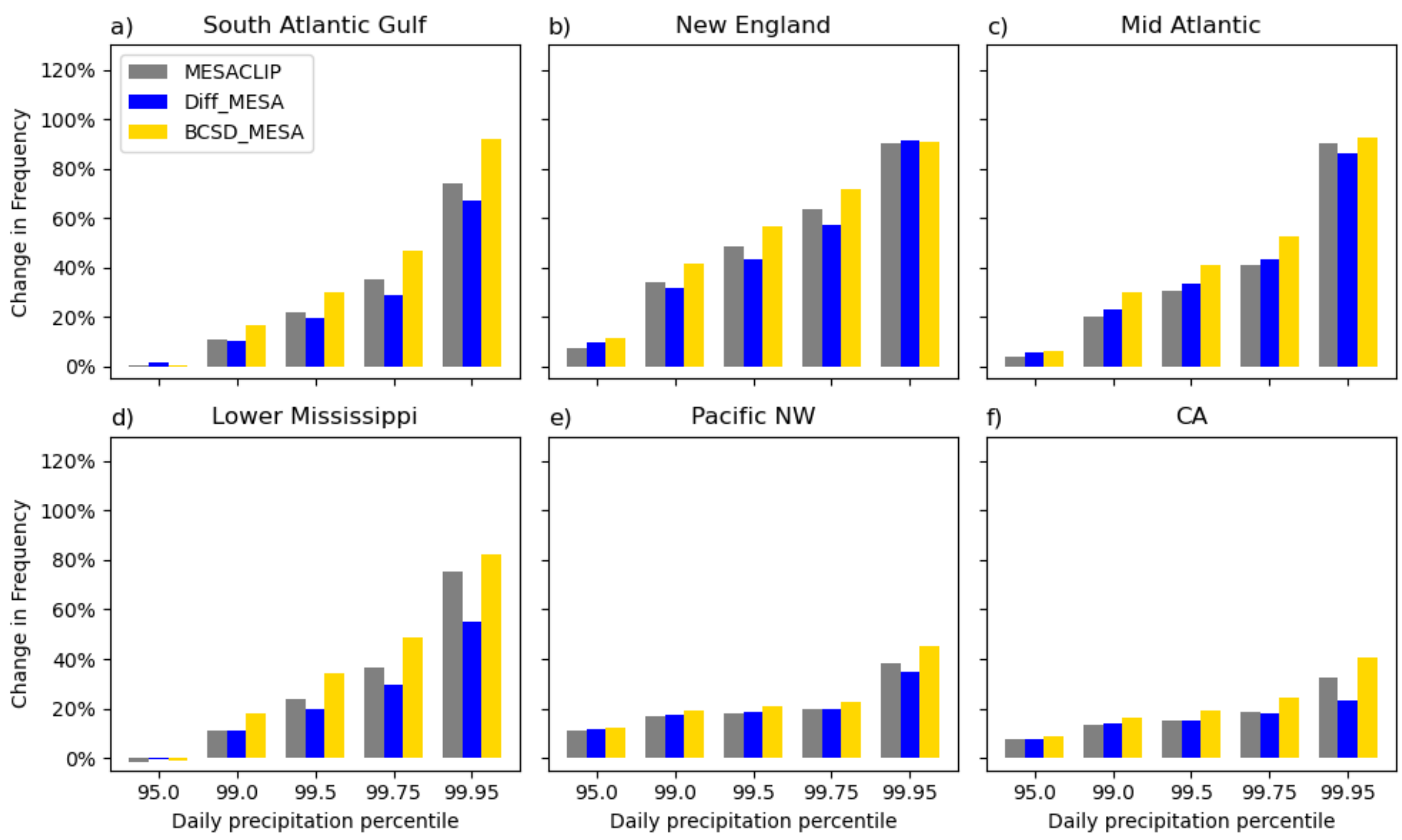

**Figure 7:** *Percent change in frequency (return rate) of various precipitation percentiles between the historical and RCP8.5 periods in the (a) South Atlantic Gulf, (b) New England, (c) Mid Atlantic, (d) Lower Mississippi, (e) Pacific Northwest, and (f) California watersheds, as projected by MESACLIP (gray), diffusion-downscaled MESACLIP (Diff_MESA, blue), and BCSD-downscaled MESACLIP (yellow). Quantiles are calculated using aggregated values from only historically climatologically wet grid cells (greater than 2 mm/day) in each watershed. The 95th, 99th, 99.5th, 99.75th, and 99.95th percentiles correspond to return periods of 20, 100, 200, 400, and 2000 days respectively.*

### *3.3 CESM2 climate change signal with and without SAI*

The diffusion downscaler faithfully reproduces the climate change signal in MESACLIP across various metrics (section 3.2), as well as reproduces historical MESACLIP precipitation extremes when conditioned on historical CESM2 coarse inputs (section 3.1). This gives confidence that the diffusion downscaler can be applied to future CESM2 climate scenarios, for which there is no “true” high-resolution precipitation with which to compare. Specifically, we now use the diffusion model to downscale future CESM2 scenarios with and without SAI, assessing how the response of precipitation extremes to climate change varies in each scenario.

Figure 8 displays the differences in mean and Rx1day precipitation under the SSP2-4.5 and ARISE-SAI scenarios (2045-2069) compared to the historical period (1980-2004), as projected by the diffusion downscaler. Results using BCSD show generally similar patterns (Figure S7). Under SSP2-4.5, the diffusion model projects the strongest mean precipitation increases in the Southeastern United States, New England, and the west coast (Figure 8a). Interestingly, the change in mean precipitation over CONUS is larger under the SSP2-4.5 scenarios than the RCP8.5 scenario in MESACLIP (Figure 5a), despite being a middle-of-the-road scenario with less warming. This could be attributable to different climate sensitivities in CESM1 (MESACLIP) and CESM2 (Gettelman et al., 2019b), differences in the circulation response to warming, and biases in the representation of precipitation in CESM2 due to a lower resolution. Under the ARISE-SAI scenario there is a much smaller change in downscaled mean precipitation (Figure 8c), with moderate increased precipitation over the east coast and the central United States, but weak drying over the Midwest and parts of California. Overall, this indicates that using SAI to limit warming has potential to mitigate changes in mean precipitation.

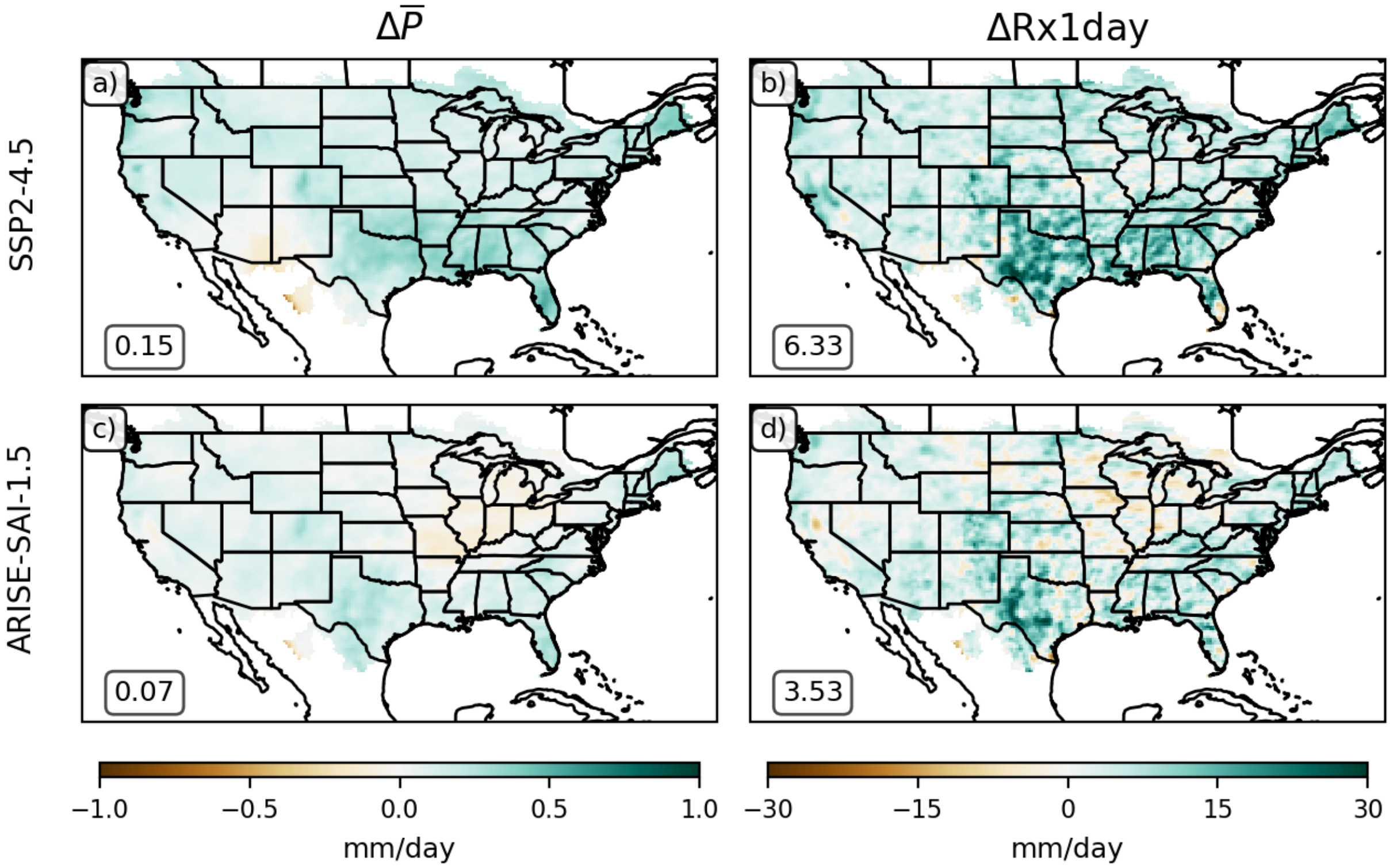

**Figure 8:** *Change in mean (left) and Rx1day (right) precipitation between the historical (1980-2004) and future (2045-2069) CESM2 periods as downscaled by diffusion model. SSP2-4.5 (top) and ARISE-SAI-1.5 (bottom). Values in bottom-left corners indicate the CONUS-average value for each subpanel.*

When analyzing changes in extreme Rx1day precipitation under SSP2-4.5, the diffusion model (Figure 8b) again projects larger mean increases in Rx1day precipitation (CONUS-averaged) compared to in the RCP8.5 scenario (Figure 5b,d). The largest increases in extreme precipitation are projected to occur in the Southeastern United States and New England. By comparison, the CONUS-averaged changes in Rx1day precipitation are nearly halved under the ARISE-SAI scenario downscaled projections (Figure 8d), indicating a potential reduction in the risks associated with extreme precipitation compared to the SSP2-4.5 scenario, generally in line with coarse ESM analyses of extremes under SAI scenarios in other regions (e.g. Jiang et al., 2024; Ji et al., 2018; Quagraine et al., 2025). The mitigation of risk varies from region to region, however, as analysis of the probability distributions of Rx1day precipitation show comparable wet shifts in the Mid Atlantic and Pacific Northwest in both future scenarios (Figure S8).

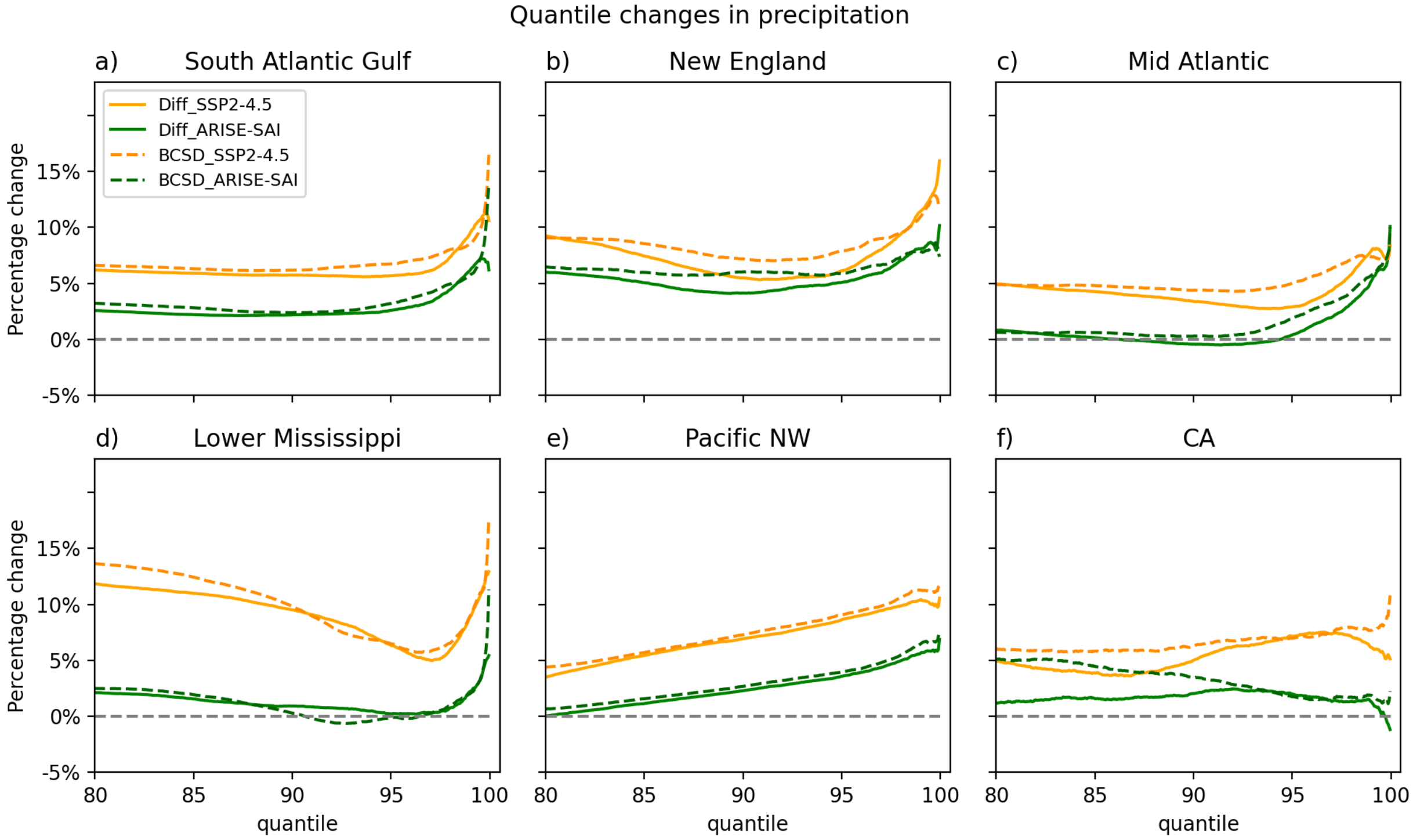

**Figure 9:** *Percent change in daily precipitation rates at various quantiles between the historical CESM2 scenario (1980-2004) and future CESM2 scenarios (2045-2069) in the (a) South Atlantic Gulf, (b) New England, (c) Mid Atlantic, (d) Lower Mississippi, (e) Pacific Northwest, and (f) California watersheds. Quantiles are calculated using aggregated values from only historically climatologically wet grid cells (greater than 2 mm/day) in each watershed. Change for SSP2-4.5 (orange) and ARISE-SAI-1.5 (green) for diffusion (solid) and BCSD (dashed).*

We repeat the quantile analysis from MESACLIP (Figure 6) for the CESM2 future scenarios (Figure 9) using the diffusion downscaled projections and BCSD. As in MESACLIP, the percentage increase in the most extreme precipitation quantiles is larger than for moderate quantiles, except in California, which provides some confidence that the diffusion downscaler is producing realistic projections. Additionally, for the most extreme precipitation percentiles in the eastern United States, the diffusion-downscaled future intensification is still generally lower than in MESACLIP, which agrees with physical expectations as less warming occurs in the CESM2 SSP2-4.5 scenarios compared to the MESACLIP RCP8.5 scenario (not shown). While the downscalers project ARISE-SAI to experience lower percentage increases than in SSP2-4.5, there are still increases of 5-10% for the 99.95th quantile across most watersheds, even when moderate precipitation quantiles do not change considerably.

Although generating mostly similar patterns, BCSD tends to result in slightly larger increases for most quantiles. The ramifications of these differences are more apparent when analyzing the change in frequency of extreme precipitation events (Figure 10), where future extreme precipitation frequency changes projected by BCSD are almost twice as large as those projected by the diffusion model in the South Atlantic Gulf, Lower Mississippi, and California. This highlights the importance of choosing an appropriate downscaling method when predicting changes in precipitation extremes in the future.

Focusing on the diffusion downscaler projections, extreme events increase in frequency by a larger amount than less extreme events (Figure 10), similar to MESACLIP (Figure 7). Across nearly all return intervals, ARISE-SAI results in a smaller increase in frequency of extreme rainfall than in SSP2-4.5. However, events exceeding the 99.95th quantile from the historical period (i.e. 2000 day return interval), still occur 30-60% more across the Eastern United States in ARISE-SAI downscaled projections, although with considerable ensemble spread. Only in California and the Lower Mississippi is the majority of the climate change signal reduced by SAI, according to the diffusion model. Overall, while there is some mitigation compared to SSP2-4.5, the downscaled projections suggest that a future under ARISE-SAI-1.5 still might experience a considerable increase in precipitation extremes compared to the historical period, at least in certain regions of the United States. This aligns with previous expectations that ARISE-SAI-1.5 still leads to a detectable climate change signal (e.g. Mamalakis et al., 2023). This is likely due to the fact that this scenario by design still includes a global temperature increase of 1.5° compared to the preindustrial period. However, there exist other SAI simulations with more aggressive temperature targets that should be a subject for future downscaling work (e.g. ARISE-SAI-1.0).

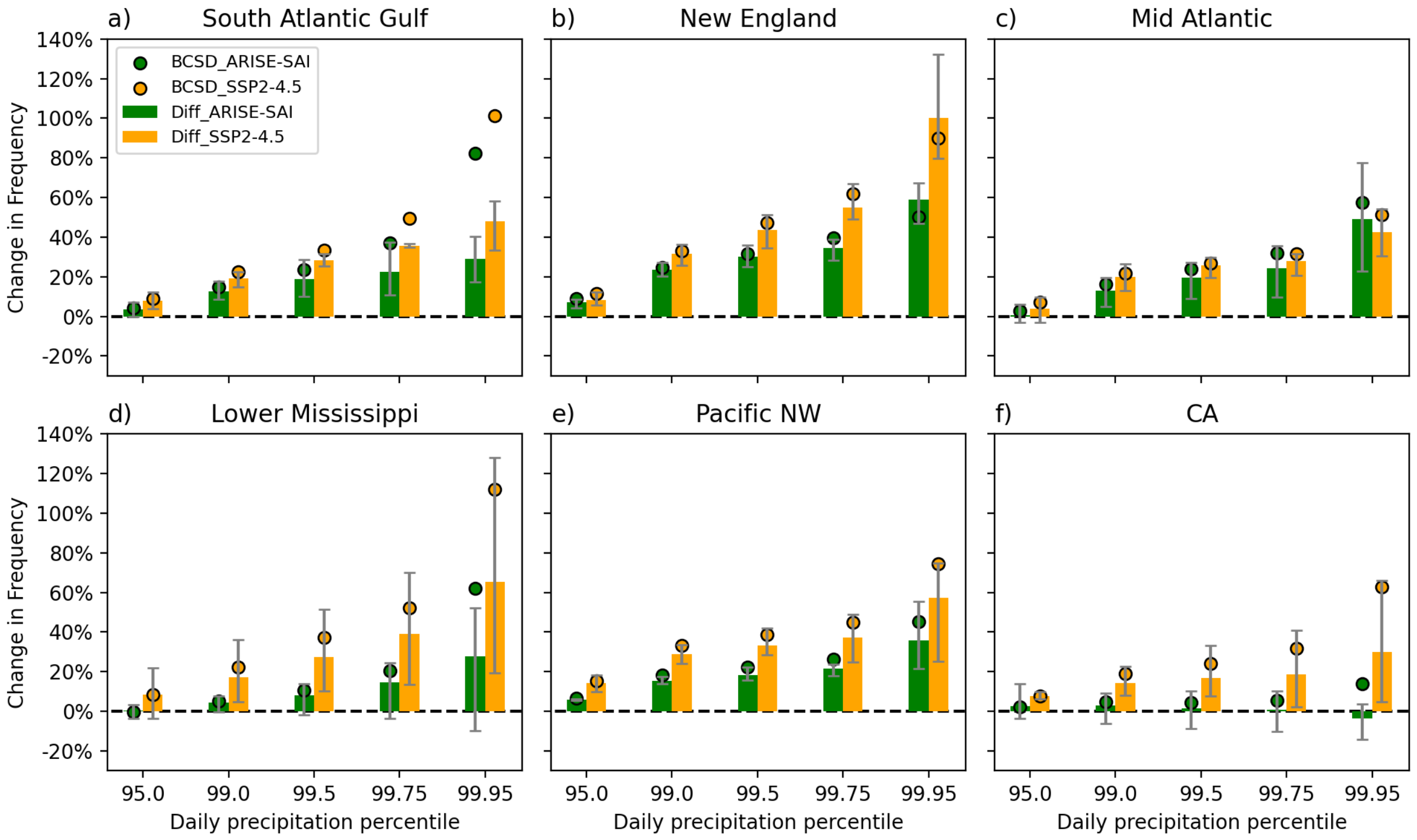

**Figure 10.** *As in Figure 7, but for the percent changes between the CESM2 historical period and the future ARISE-SAI-1.5 scenario (green) and the future SSP2-4.5 scenario (orange). All values derived from diffusion-downscaled precipitation of each scenario. Diffusion downscaling (bar) and BCSD (single dots). Error bars represent the ensemble range.*

## 4. Discussion

We use a diffusion framework to train a deep learning neural network to produce daily high-resolution (0.25°) precipitation fields over CONUS given coarse (4°) daily precipitation, total column water, and 500 hPa geopotential height. The purpose of this neural network downscaler is to improve projections of fine-scale extreme weather in future climate scenario simulations, which typically rely on coarse-resolution (>=1°) ESM's due to computational constraints. In particular, we apply the diffusion downscaler to future coarse CESM2 climate simulations with and without stratospheric aerosol injection, a potential climate intervention strategy that could mitigate global-mean warming, but has uncertain consequences and risks on regional scales.

The downscaler is trained using 750 years of historical and RCP8.5 high-resolution climate simulations from MESACLIP (Chang et al., 2020). A benefit of training on MESACLIP, rather than observations or reanalysis (which are available at similar 0.25 °resolution), is the ability to train the neural network on future warmer climates, which is critical to increase their generalization capability for projecting climate change signals when applied to unseen future scenarios (Rampal et al., 2024b). It is also reasonable to expect that the model can generalize well to SAI scenarios, as the coarse conditioning variables implicitly capture the impacts of SAI. We assume the relationship between fine-scale precipitation and the coarse conditioning variables is similar in both SAI and non-SAI worlds, although it would be beneficial to verify this in future work, which requires expanding the limited amount of high-resolution SAI simulations (e.g. Feder et al., 2026).

When validating the performance of the neural network to downscale the historical climate, it considerably outperforms a traditional statistical downscaling baseline (BCSD), better capturing the full distribution of precipitation values. The improvement is especially noticeable for precipitation extremes (e.g. 99.95th percentile, Rx1day), which are significantly underestimated by BCSD but accurately reproduced by the neural network. Most importantly, the diffusion downscaler successfully reproduces historical MESACLIP precipitation variability when using coarse predictors from either MESACLIP test data or out-of-distribution CESM2 historical simulations. This suggests that CESM2 and MESACLIP contain sufficiently similar large-scale atmospheric variability, which allows the neural network to generalize to CESM2 inputs using learned relationships between coarse conditioning and fine-scale precipitation from MESACLIP.

The diffusion downscaler also closely matches the climate change signal from MESACLIP, capturing the spatial pattern and magnitude of shifts in both mean and Rx1day precipitation across CONUS. This includes accurately representing the faster intensification of precipitation extremes compared to moderate precipitation, as simulated by MESACLIP and theorized from dynamical principles (O'Gorman and Schneider 2009). Additionally, because the diffusion downscaler realistically represents fine-scale variability and stochasticity, it can provide an improved measurement of internal variability and uncertainty at small spatial scales, critical when assessing risks and probabilities of extreme weather in future climate projections.

It is not strictly possible to test the performance of the diffusion downscaler on capturing the climate change signal in out-of-distribution coarse ESM's for which there is no high-resolution counterpart. However, the ability of the neural network to generalize during the historical period to CESM2 inputs, as well as the ability to capture the full climate change signal in MESACLIP, gives confidence to downscale coarse CESM2 future climate simulations under an SSP2-4.5 scenario and a paired SAI scenario that limits warming to 1.5 degrees over preindustrial.

Within the diffusion downscaling framework, we find that the ARISE-SAI scenario (Richter et al., 2022) leads to reduced changes in both mean and Rx1day precipitation over CONUS compared to an SSP2-4.5 scenario without SAI. On a CONUS-wide average, the downscaled projections suggest nearly a 45% reduction in Rx1day intensification in the SAI scenario compared to the non-SAI scenario. However, this mitigation varies regionally. While the non-SAI scenario generally leads to larger intensification of downscaled extreme precipitation, the ARISE-SAI scenario still displays the majority of the non-SAI signal in certain regions like the Mid Atlantic and Pacific Northwest, which are characterized by comparable increases in the frequency of the most extreme historical precipitation quantile (99.95th percentile). Internal variability likely plays a considerable role in regional differences and uncertainties. BCSD produces frequency changes that are vastly largely, which highlights the potential added value of the diffusion downscaler over traditional downscaling methods that have been used to project SAI impacts (Jiang et al., 2024; Ji et al., 2018; Quagraine et al., 2025). Our results suggest that in this SAI scenario, changes in extreme precipitation may be generally reduced compared to the non-SAI scenario, but could still be distinguishable from the reference period (Mamalakis et al., 2023).

Overall, machine learning methods, in particular generative methods such as diffusion, provide a way to skillfully downscale coarse ESM outputs by generating both realistic samples and reproducing important climate statistics, all while learning in an entirely data-driven manner. However, there are some limitations and areas for future improvement, especially when applying the learned neural network to unseen ESM's. Quantile delta mapping as used here is able to correct temporal distributions of the coarse ESM inputs, but not the temporal coherence, spatial variability, or intervariable relationships, which results in the slightly different climate statistics over the historical period for MESACLIP or CESM2 conditioning. More advanced bias-correction techniques, including machine learning methods such as rectified flows (e.g. Wan et al., 2023; Wan et al., 2026), could improve the biases in the CESM2 input and thus further improve the downscaled projections of extreme precipitation. Additionally, this study is limited to daily variability; however, some impacts of extreme precipitation (such as landslides and flooding), are significantly affected by the temporal evolution of events that last multiple days. Bias correction methods and downscaling methods that handle temporal coherence (e.g. video diffusion, Srivastava et al., 2023), are an important avenue to research to analyze such impacts. While applied here only to a small ensemble of CESM2 SAI and non-SAI scenarios, a main benefit of the diffusion downscaler (and most other machine learning downscalers), is the cheap computational expense. As such, application to larger ensembles of coarse ESM simulations and a wider variety of climate intervention scenarios would allow for a more nuanced and thorough analysis of the uncertainties of future projected changes.